\documentclass[10pt, conference]{IEEEtran}
\IEEEoverridecommandlockouts
\usepackage[letterpaper, left=0.65in, right=0.65in, bottom=1.03in, top=0.75in]{geometry}
\usepackage{cite}
\usepackage{amsmath,amssymb,amsfonts}
\usepackage{graphicx}
\usepackage[caption=false,font=footnotesize]{subfig}
\usepackage{textcomp}
\usepackage{xcolor}

\usepackage{tabularx}
\usepackage{multirow}
\usepackage{diagbox}
\usepackage[binary-units]{siunitx} 
\DeclareSIUnit{\bps}{bps}
\usepackage{flushend}
\usepackage{amsthm}

\begin{document}

\title{Cost-Effectiveness Analysis and Design of Cost-Efficient Cell-Free Massive MIMO Systems}

\author{\IEEEauthorblockN{Wei Jiang\IEEEauthorrefmark{1} and Hans D. Schotten\IEEEauthorrefmark{2}}
\IEEEauthorblockA{\IEEEauthorrefmark{1}German Research Center for Artificial Intelligence (DFKI)\\
  }
\IEEEauthorblockA{\IEEEauthorrefmark{2}Department of Electrical and Computer Engineering, University of Kaiserslautern  (RPTU)\\
 }
}
\maketitle

\begin{abstract}
Cell-free massive multi-input multi-output (MIMO) has recently attracted much attention, attributed to its potential to deliver uniform service quality. However, the adoption of a cell-free architecture raises concerns about the high implementation costs associated with deploying numerous distributed access points (APs) and the need for fronthaul network installation. To ensure the sustainability of next-generation wireless networks, it is crucial to improve cost-effectiveness, alongside achieving high performance. To address this, we conduct a cost analysis of cell-free massive MIMO and build a unified model with varying numbers of antennas per AP. Our objective is to explore whether employing multi-antenna APs could reduce system costs while maintaining performance. The analysis and evaluation result in the identification of a cost-effective design for cell-free massive MIMO, providing valuable insights for practical implementation.
\end{abstract}
\section{Introduction}

In a traditional cellular network \cite{Ref_jiang2024TextBook}, a base station (BS) is positioned at the center of a cell within a network of cells. High quality of service (QoS) is delivered to users at the cell center, close to the BS. However, the users at the cell edge experience worse QoS due to considerable distance-dependent path loss, strong inter-cell interference, and inherent handover issues within the cellular architecture. The performance gap between the cell center and edge is not merely a minor concern; it is substantial \cite{Ref_jiang2023celledge}. 
Recently, cell-free massive multi-input multi-output (MIMO) \cite{Ref_ngo2017cellfree} has garnered much attention in both academia and industry due to its high potential for the upcoming sixth-generation (6G) systems \cite{Ref_jiang2021road}. There are no cells or cell boundaries. Instead, a multitude of distributed low-power, low-cost access points (APs) simultaneously serve users over the same time-frequency resource \cite{Ref_nayebi2017precoding}. It perfectly matches some 6G scenarios, such as private or campus networks, with dedicated coverage areas like factories, stadiums, shopping malls, airports, railway stations,  and exhibition halls. The cell-free architecture ensures uniform QoS for all users, effectively addressing the issue of under-served areas commonly encountered at the edges of conventional cellular networks \cite{Ref_buzzi2020usercentric}. 

Despite its considerable potential, the adoption of a cell-free architecture poses challenges related to high implementation costs associated with deploying numerous distributed APs and the necessity of installing a large-scale fronthaul network \cite{Ref_masoumi2020performance}. Deploying a traditional wireless network is already time-consuming and costly mainly due to acquiring and maintaining base station sites. In the cell-free system, this challenge intensifies as hundreds of suitable sites must be identified within a small area for wireless AP installations. Moreover, the installation of a large-scale fiber-cable network to interconnect these APs raises the expenditure and energy consumption \cite{Ref_jiang2024hierarchical}. To ensure the sustainability of next-generation wireless networks, improving the \textit{cost-effectiveness} of cell-free systems is crucial, alongside achieving high performance. 

To tackle this issue, we perform a cost analysis on cell-free massive MIMO, acknowledging that the overall cost including capital expenditure (CaPEX) and operational expenditure (OPEX) depends on the number of distributed wireless sites. Notably, the utilization of additional antennas at each AP does not induce extra site acquisition, and fiber connections, or incur supplementary maintenance costs. Consequently, reducing the number of APs by mounting multiple antennas on each AP can reduce implementation costs. However, a reduction in AP density compromises per-user spectral efficiency and sum capacity. This prompts us to seek a balance between performance and cost by determining the optimal number of antennas per AP. To provide a quantitative assessment for designing a cost-efficient architecture, a unified model encompassing cell-free architectures with varying antenna numbers per AP is built. We analyze and evaluate the spectral efficiency and cost-effectiveness in both downlink and uplink scenarios with maximal-ratio or zero-forcing schemes. 


\section{Cell-Free Massive MIMO System}

In this paper, we argue that the cost-effectiveness of a cell-free massive MIMO system is proportionally related to the number of distributed AP sites. This is due to the fact that acquiring and maintaining AP sites incurs significant costs, and the installation of a large-scale fiber-cable network to interconnect these APs amplifies the expenses. However, the fees associated with adding extra antennas to each AP are marginal since a unique advantage for massive MIMO systems is the use of low-cost antennas and RF components \cite{Ref_marzetta2015massive}. 
\subsubsection{System Model}
\begin{figure}[!tpbh]
    \centering
    \includegraphics[width=0.43\textwidth]{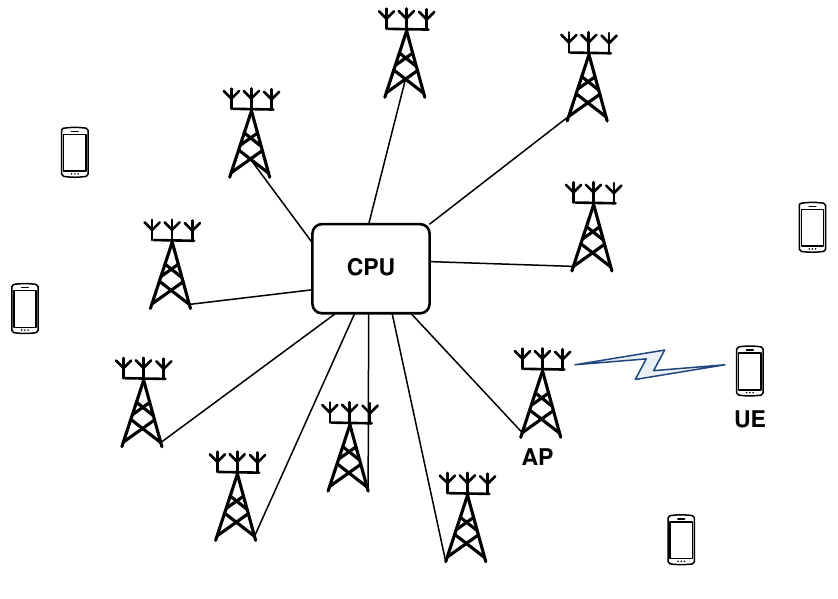}
    \caption{A unified model for cell-free massive MIMO with a varying number of antennas per AP, where a total of $M$ antennas are distributed over $N_{AP}$ sites. If $N_{AP}=M$, it stands for typical cell-free massive MIMO with single-antenna APs, whereas we are interested in applying multi-antenna APs to improve cost-effectiveness by reducing the number of AP sites.  }
    \label{fig:SystemModel}
\end{figure}

To offer a quantitative analysis for designing a cost-effective architecture, we formulate a general model for cell-free massive MIMO, allowing for the flexibility to set the number of antennas per AP. A total number of $M$ service antennas are distributed across $N_{AP}$ sites, where $1< N_{AP} \leqslant M$. Each AP is equipped with $N_{t}$ antennas, adhering to $N_t\times N_{AP}=M$. These APs serve a few single-antenna user equipments (UEs) within a designated coverage area and the number of users $K\ll M$. In contrast to previous studies, our model accommodates a \textit{variable} number of antennas per AP, covering conventional cell-free massive MIMO scenarios with $M$ distributed single-antenna APs ($N_{t}=1$), like \cite{Ref_ngo2017cellfree, Ref_ngo2018total, Ref_nayebi2017precoding} and multi-antenna APs ($N_{t}>1$) \cite{Ref_buzzi2017cellfree}. As depicted in \figurename \ref{fig:SystemModel}, a central processing unit (CPU) coordinates all APs through a fronthaul network to simultaneously serve all users over the same time-frequency resource.
\subsubsection{Channel Model}
The channel coefficient connecting antenna $m$ (for all $m=1,\ldots,M$) to UE $k$ (for all $k=1,\ldots,K$) is represented as a circularly symmetric complex Gaussian random variable, denoted as $g_{mk}\in \mathcal{CN}(0,\beta_{mk})$. Here, $\beta_{mk}$ denotes large-scale fading, including path loss and shadowing. To alleviate the considerable downlink pilot overhead scaling with the number of service antennas, time-division duplex (TDD) is employed to separate the downlink and uplink signal transmission with the assumption of perfect channel reciprocity. Hence, each coherent interval is divided into three phases: uplink training, uplink data transmission, and downlink data transmission \cite{Ref_nayebi2017precoding}. During the uplink training phase, UEs transmit orthogonal pilot sequences to acquire instantaneous channel state information (CSI). Using minimum mean-square error (MMSE) estimation \cite{Ref_ngo2017cellfree}, the network obtains channel estimates $\hat{g}_{mk}$, following a complex normal distribution $\mathcal{CN}(0,\alpha_{mk})$ with $\alpha_{mk}=\frac{p_u\beta_{mk}^2}{p_u \beta_{mk} + \sigma_n^2}$, where $p_u$ represents the power of the UE transmitter, and $\sigma_n^2$ is the variance of noise \cite{Ref_jiang2021cellfree}. The estimation error is defined as $\tilde{g}_{mk} = g_{mk} - \hat{g}_{mk}$, following the distribution of $\mathcal{CN}(0,\beta_{mk}-\alpha_{mk})$. In the case of multi-antenna APs, the large-scale fading between user $k$ and any antenna of the same AP $q$ (for $q=1,\ldots, N_{AP}$) is identical. This assumption leads to $\beta_{mk}=\beta_{qk}$ and $\alpha_{mk}=\alpha_{qk}$, where $ m\in \{(q-1)N_t{+}1,(q-1)N_t{+}2,\ldots,qN_t\}$.

\section{Spectral-Efficiency and Cost-Effectiveness Analysis in Uplink Transmission}
During uplink data transmission, each of the UEs aims to deliver its unit-variance, independent information-bearing symbol $x_k$ simultaneously to the APs, with UE $k$ transmitting $\sqrt{\eta_k}x_k$. The covariance matrix of the transmit vector $\textbf{x}=[x_1,\ldots,x_K]^T$ adheres to $\mathbb{E}[\textbf{x}\textbf{x}^H]=\mathbf{I}_K$, where $\mathbf{I}_K$ is the identity matrix of size $K$. The power coefficient is constrained by $0\leqslant \eta_k \leqslant 1$. The received signals at AP $q$ can be expressed as an $N_t\times 1$ vector: 
\begin{equation}  
    \textbf{y}_q = \sqrt{p_u} \sum_{k=1}^K\textbf{g}_{qk} \sqrt{\eta_k}x_k + \textbf{n}_q,
\end{equation}
where $\textbf{g}_{qk}$ denotes an $N_t\times 1$ channel signature between user $k$ and AP $q$ and the receiver noise $\textbf{n}_q=[n_1,\ldots,n_{N_t}]^T$ follows a complex normal distribution $\mathcal{CN}(\mathbf{0},\sigma^2_n\mathbf{I}_{N_t})$. 

In contrast to maximum likelihood, linear detection stands out for its lower complexity while still achieving commendable performance. Maximum-ratio combining (MRC), also known as matched filtering, is commonly employed for uplink detection in massive MIMO \cite{Ref_jiang20236GCH9}. The underlying principle is to maximize the strength of the desired signal while disregarding inter-user interference (IUI) \cite{Ref_jiang2022opportunistic}. To detect the symbols, AP $q$ multiplies the received signals with the conjugate of its locally obtained channel estimates represented by $\hat{\textbf{g}}_{qk}$, consisting of $\hat{g}_{mk}$, $\forall m\in \{(q-1)N_t{+}1,(q-1)N_t{+}2,\ldots,qN_t\}$. Then, the result $\hat{\textbf{g}}_{qk}^*\otimes \textbf{y}_q$, where $\otimes$ marks the Hadamard (element-wise) product, is delivered to the CPU via the fronthaual network.

As a result, the CPU observes
\begin{align}
    \label{eqnUPLINKmodel}  \nonumber
    \textbf{y} &= \hat{\textbf{G}}^H \left( \sqrt{p_u} \textbf{G} \textbf{D}_\eta \textbf{x} + \textbf{n}  \right)\\
    &= \hat{\textbf{G}}^H \left(\sqrt{p_u} \sum_{k=1}^K\textbf{g}_k \sqrt{\eta_k}x_k + \textbf{n}\right).
\end{align} 
Here, $\textbf{G}$ stands for an $M\times K$ channel matrix with the $(m,k)^{th}$ entry of $g_{mk}$, namely $\left[\textbf{G}\right]_{mk}=g_{mk}$, $\hat{\textbf{G}}$ is the matrix of channel estimates, i.e., $[\hat{\textbf{G}}]_{mk}=\hat{g}_{mk}$, $\textbf{g}_k$ represents the channel signature for user $k$ or the $k^{th}$ column of $\textbf{G}$, and $\hat{\textbf{g}}_k$ denotes the $k^{th}$ column of $\hat{\textbf{G}}$. Additionally, $\textbf{D}_\eta$ is a diagonal matrix given by $\textbf{D}_\eta = \mathrm{diag}([\eta_1,\ldots,\eta_K])$, and the receiver noise $\textbf{n}=[n_1,\ldots,n_M]^T$ follows a complex normal distribution $\mathcal{CN}(\mathbf{0},\sigma^2_n\mathbf{I}_M)$. 
Decomposing \eqref{eqnUPLINKmodel} yields the received signal for detecting $x_k$ as 
\begin{align}  \nonumber \label{Eqn_UlCBFuserK}
    y_k &= \hat{\textbf{g}}_k^H \left(  \sqrt{p_u} \sum_{k=1}^K\textbf{g}_k \sqrt{\eta_k}x_k {+} \textbf{n} \right)\\
    &= \sqrt{p_u \eta_k} \hat{\textbf{g}}_k^H \textbf{g}_k  x_k  + \sqrt{p_u}\sum_{i=1,i\neq k}^K \hat{\textbf{g}}_k^H \textbf{g}_{i} \sqrt{\eta_{i}} x_{i}+\hat{\textbf{g}}_k^H\textbf{n}.
\end{align} 
Considering the vector of channel estimate errors denoted as $\tilde{\textbf{g}}_k=\textbf{g}_k-\hat{\textbf{g}}_k$, \eqref{Eqn_UlCBFuserK} is further derived as
\begin{align}  \nonumber \label{Eqn_Xjakodfj}
    y_k &= \sqrt{p_u \eta_k} \hat{\textbf{g}}_k^H (\hat{\textbf{g}}_k + \tilde{\textbf{g}}_k ) x_k  + \sqrt{p_u}\sum_{i=1,i\neq k}^K \hat{\textbf{g}}_k^H \textbf{g}_{i} \sqrt{\eta_{i}} x_{i}+\hat{\textbf{g}}_k^H\textbf{n}\\ \nonumber
    &= \sqrt{p_u \eta_k} \|\hat{\textbf{g}}_k\|^2  x_k  + \sqrt{p_u \eta_k}  \hat{\textbf{g}}_k^H \tilde{\textbf{g}}_k  x_k \\ &+\sqrt{p_u}\sum_{i=1,i\neq k}^K \hat{\textbf{g}}_k^H \textbf{g}_{i} \sqrt{\eta_{i}} x_{i}+\hat{\textbf{g}}_k^H\textbf{n}.
\end{align} 

The CPU gets the full CSI exclusively when the CSI obtained from all APs is transmitted via the fronthaul network, or if the CPU conducts centralized estimation using observations also provided by the APs. Given the high signaling overhead, it is sensible to assume that the CPU is only equipped with knowledge of the channel statistics \cite{Ref_nayebi2017precoding, Ref_ngo2017cellfree}. That is to say, the CPU detects the received signals based on $\alpha_{mk}$, rather than $\hat{g}_{mk}$, $\forall m,k$.
Thus, \eqref{Eqn_Xjakodfj} is rewritten as
\begin{align}  \nonumber \label{eqn_cfmULdetection}
    y_k &= \underbrace{\sqrt{p_u \eta_k} \mathbb{E}\left[\left\|\hat{\textbf{g}}_k\right\|^2\right]  x_k }_{\mathcal{S}_0:\:desired\:signal}+\underbrace{\sqrt{p_u \eta_k} \left(\|\hat{\textbf{g}}_k\|^2 {-}\mathbb{E}\left[\|\hat{\textbf{g}}_k\|^2\right]\right)  x_k}_{\mathcal{I}_4:\:channel\:uncertainty\:error} \\ \nonumber
    &+ \underbrace{\sqrt{p_u \eta_k}  \hat{\textbf{g}}_k^H \tilde{\textbf{g}}_k  x_k}_{\mathcal{I}_1:\:channel\:estimation\:error} +\underbrace{\sqrt{p_u}\sum_{i=1,i\neq k}^K \hat{\textbf{g}}_k^H \textbf{g}_{i} \sqrt{\eta_{i}} x_{i}}_{\mathcal{I}_2:\:inter-user\:interference}\\
    &+\underbrace{\hat{\textbf{g}}_k^H\textbf{n}}_{\mathcal{I}_3:\:noise}.
\end{align}

\paragraph{Spectral-Efficiency Analysis} The terms $\mathcal{S}_0$, $\mathcal{I}_1$, $\mathcal{I}_2$, $\mathcal{I}_3$, and $\mathcal{I}_4$ as defined in \eqref{eqn_cfmULdetection} exhibit mutual uncorrelation. As stated in \cite{Ref_hassibi2003howmuch}, the worst-case noise for mutual information corresponds to Gaussian additive noise with a variance equal to the sum of the variances of $\mathcal{I}_1$, $\mathcal{I}_2$, $\mathcal{I}_3$, and $\mathcal{I}_4$.
Thus, the uplink achievable rate for user $k$ is lower bounded by $R_k= \log(1+\gamma_{k}^{ul})$,
where
\begin{align}  \label{cfmmimo:formularSNR} \nonumber
    \gamma_{k}^{ul}  &= \frac{\mathbb{E}\left[|\mathcal{S}_0|^2\right]}{\mathbb{E}\left[|\mathcal{I}_1+\mathcal{I}_2+\mathcal{I}_3+\mathcal{I}_4|^2\right]}\\
         &= \frac{\mathbb{E}\left[|\mathcal{S}_0|^2\right]}{\mathbb{E}\left[|\mathcal{I}_1|^2\right]+\mathbb{E}\left[|\mathcal{I}_2|^2\right]+\mathbb{E}\left[|\mathcal{I}_3|^2\right]+\mathbb{E}\left[|\mathcal{I}_4|^2\right]}
\end{align}
with 
\begin{align}  \label{APPEQ1}
    \mathbb{E}\left[|\mathcal{S}_0|^2\right] & = p_u\eta_k N_t^2\left( \sum_{q=1}^{N_{AP}} \alpha_{qk} \right)^2\\ \label{APPEQ2}
    \mathbb{E}\left[|\mathcal{I}_1|^2\right] & = p_u\eta_kN_t\sum_{q=1}^{N_{AP}} (\beta_{qk}-\alpha_{qk})\alpha_{qk}\\ \label{APPEQ3}
    \mathbb{E}\left[|\mathcal{I}_2|^2\right] & = p_u N_t \sum_{i=1,i\neq k}^K \eta_i \sum_{q=1}^{N_{AP}}  \beta_{qi} \alpha_{qk}\\  \label{APPEQ4}
    \mathbb{E}\left[|\mathcal{I}_3|^2\right] & = \sigma_n^2N_t\sum_{q=1}^{N_{AP}} \alpha_{qk}\\
    \mathbb{E}\left[|\mathcal{I}_4|^2\right] & = N_t \sum_{q=1}^{N_{AP}} \alpha_{qk}^2.
\end{align}
Substituting the above terms into \eqref{cfmmimo:formularSNR}, yields 
\begin{equation} \label{eqn:SNR_UL_CBF_Cstats}
    \gamma_{k}^{ul} =  \frac{p_u \eta_k N_t^2\left(\sum_{q=1}^{N_{AP}}  \alpha_{qk}  \right)^2}
    {p_u N_t \sum_{i=1}^{K}  \eta_{i}  \sum_{q=1}^{N_{AP}}  \alpha_{qk}\beta_{qi} +\sigma^2_nN_t \sum_{q=1}^{N_{AP}}  \alpha_{qk}    }.
\end{equation}
By now, we get the closed-form expression of uplink per-user spectral efficiency for a generalized system with $1< N_{AP} \leqslant M$ single- or multi-antenna APs.

\paragraph{Cost Analysis} In evaluating the cost-effectiveness of the system, the overall cost of a cell-free massive MIMO system is formulated as
\begin{align} \label{EQN_CostFormular} \nonumber
    C &= c_{ls}+ c_{cpu}+ c_{mo}\\ 
      &+ N_{AP}\Bigl[ c_{sc}+ c_{ps}+ c_{fb}+ c_{bb}+ N_{t} (c_{ant}+ c_{rf})\Bigr].
\end{align}
This comprehensive cost includes various components such as $c_{ls}$ representing fees associated with acquiring spectrum licenses from regulatory authorities, $c_{cpu}$ denoting the expense of constructing the CPU, and $c_{mo}$ covering the costs of electricity consumption, routine maintenance, network management, monitoring, and insurance. Furthermore, $c_{sc}$ accounts for the expenses related to obtaining or leasing space for a wireless site and site construction, $c_{fb}$ addresses the establishment cost of a fiber optic connection to the CPU,  $c_{ps}$ encompasses the installation cost of power supply infrastructure, $c_{bb}$ represents the cost of a baseband unit, while $c_{ant}$ and $c_{rf}$ correspond to the cost per antenna and the cost of the RF chain, respectively. 

To investigate the influence of the number of antennas per AP on cost-effectiveness, we redefine \eqref{EQN_CostFormular} in a new form as:
\begin{equation}
    C= N_{AP}\Bigl[ C_{f}+ N_{t} C_v\Bigr].
\end{equation}
Here, the combined costs of $c_{ls}+ c_{cpu}+ c_{mo}$ are equally allocated to each AP by dividing by $N_{AP}$, $C_f$ denotes the fixed cost independent of the number of AP antennas 
\begin{equation}
    C_f= \frac{c_{ls}+ c_{cpu}+ c_{mo}}{N_{AP}}+  c_{sc}+ c_{ps}+ c_{fb}+ c_{bb},
\end{equation}
and the fees associated with the number of AP antennas are represented by $C_v=c_{ant}+ c_{rf}$. Consequently, the cost-effectiveness of the system in the uplink can be assessed through the ratio between the sum rate and the overall cost, i.e.,
\begin{equation} \label{qneddd}
    \Gamma_{ul}=\frac{\sum_{k=1}^K \log(1+\gamma_{k}^{ul})}{N_{AP}( C_{f}+ N_{t} C_v )}.
\end{equation}

\section{Spectral-Efficiency and Cost-Effectiveness Analysis in Downlink Transmission}
In cell-free massive MIMO systems, the spatial multiplexing of information symbols is commonly realized using two linear precoding techniques: conjugate beamforming (CBF) \cite{Ref_ngo2017cellfree} and zero-forcing precoding (ZFP) \cite{Ref_jiang2021impactcellfree}.
The symbols intended for $K$ users is denoted by $\textbf{u}=[u_1,\ldots,u_K]^T$, where  $\mathbb{E}[\textbf{u}\textbf{u}^H]=\mathbf{I}_K$. Let $\textbf{B}$ represent the $M\times K$ precoding matrix with elements $\left[\textbf{B}\right]_{mk}=\sqrt{\eta_{mk}}b_{mk}$, where $\eta_{mk}$ denotes the power coefficient for the $k^{th}$ user at antenna $m$, and $b_{mk}$ is the precoding coefficient.   Adhering to the per-antenna power constraint $p_d$ and a noise vector $\textbf{w}=[w_1,\ldots,w_K]^T\in \mathcal{CN}(\mathbf{0},\sigma^2_n\mathbf{I}_K)$, the collective representation of received symbols for all users, namely $\textbf{r}=[r_1,\ldots,r_K]^T$, is given by:
\begin{equation}
    \label{eqn:RxSignal}\mathbf{r} =\sqrt{p_d} \mathbf{G}^T\mathbf{B}\mathbf{u}+\mathbf{w}.
\end{equation}
Equivalently, the $k^{th}$ user has the observation of
\begin{align}\nonumber \label{EQN_downlinkModel}
    r_k &= \sqrt{p_d} \mathbf{g}_k^T\mathbf{B}\mathbf{u}+w_k\\ 
    &= \sqrt{p_d} \mathbf{g}_k^T \sum_{i=1}^K\mathbf{b}_i u_i+w_k\\\nonumber
    &= \sqrt{p_d}\textbf{g}_k^T\textbf{b}_k u_k + \sqrt{p_d}\sum_{i=1,i\neq k}^K \textbf{g}_k^T \textbf{b}_{i}  u_{i}+w_k,
\end{align}
where $\textbf{b}_k\in \mathbb{C}^{M\times 1}$ is the $k^{th}$ column of $\textbf{B}$.

\subsubsection{Conjugate Beamforming}
Similar to the MRC technique employed in the uplink, CBF is designed to optimize the reception of the desired signal \cite{Ref_ngo2017cellfree}, aiming to maximize its strength. Its precoding matrix is given by $\left[\textbf{B}\right]_{mk}=\sqrt{\eta_{mk}}\hat{g}^*_{mk}$. 
Applying CBF, \eqref{EQN_downlinkModel} can be rewritten into an element-wise form as
\begin{align}\nonumber  \label{EQN_DLRxSignal}
    r_k  &= \sqrt{p_d} \sum_{m=1}^M \sqrt{\eta_{mk}} g_{mk}\hat{g}_{mk}^* u_k \\
         &+\sqrt{p_d}\sum_{i=1,i\neq k}^K  \sum_{m=1}^M \sqrt{\eta_{mi}} g_{mk}\hat{g}_{mi}^*  u_i  + w_k.
\end{align}
Each user only knows channel statistics $ \mathbb{E} \left[ \left |  \hat{g}_{mk} \right | ^2\right]=\alpha_{mk}$ rather than channel estimate $\hat{g}_{mk}$ since there are no downlink pilots. As a result, each user detects the signals based on channel statistics. Like \eqref{eqn_cfmULdetection}, \eqref{EQN_DLRxSignal} is transformed to
\begin{align}\nonumber 
    r_k  &= \sqrt{p_d} \sum_{m=1}^M \sqrt{\eta_{mk}} \mathbb{E} [  |  \hat{g}_{mk}  | ^2] u_k \\ \nonumber
    &+ \sqrt{p_d} \sum_{m=1}^M \sqrt{\eta_{mk}}\left( | \hat{g}_{mk}|^2-\mathbb{E} [  |  \hat{g}_{mk}  | ^2]\right) u_k \\ \nonumber
    &+ \sqrt{p_d} \sum_{m=1}^M \sqrt{\eta_{mk}} \tilde{g}_{mk}\hat{g}_{mk}^* u_k \\ 
         &+\sqrt{p_d}\sum_{i=1,i\neq k}^K  \sum_{m=1}^M \sqrt{\eta_{mi}} g_{mk}\hat{g}_{mi}^*  u_i  + w_k.
\end{align}

The power control is generally decided by site-specific large-scale fading coefficient $\beta_{qk}$, we have $\eta_{mk}=\eta_{qk}$ for all $ m\in \{(q-1)N_t{+}1,(q-1)N_t{+}2,\ldots,qN_t\}$. Applying analogous manipulations to the derivation of \eqref{eqn:SNR_UL_CBF_Cstats}, we obtain the effective SINR as
\begin{equation}  \label{EQN_DLCBFPerUserRate}
    \gamma_{k}^{cbf}=  \frac{p_d N_t^2\left(\sum_{q=1}^{N_{AP}} \sqrt{\eta_{qk}} \alpha_{qk}  \right)^2}
    {\sigma^2_n+p_d N_t \sum_{q=1}^{N_{AP}} \beta_{qk} \sum_{i=1}^{K}  \eta_{qi} \alpha_{qi} }.
\end{equation}
Let $N_{AP}=M$ and $N_t=1$,  \eqref{EQN_DLCBFPerUserRate} reverts to equation (27) in \cite{Ref_ngo2017cellfree}, illustrating the performance of conventional cell-free massive MIMO with single-antenna APs. Nevertheless, we progress beyond by covering the scenario of multi-antenna APs when $N_t>1$.
Like \eqref{qneddd}, the cost-effectiveness of CBF-based cell-free massive MIMO in the downlink can be evaluated by
\begin{equation}
    \Gamma_{dl}^{cbf}=\frac{\sum_{k=1}^K \log(1+\gamma_{k}^{cbf})}{ N_{AP}( C_{f}+ N_{t} C_v )  }.
\end{equation}

\begin{figure*}[!tbph]
\centerline{
\subfloat[]{
\includegraphics[width=0.32\textwidth]{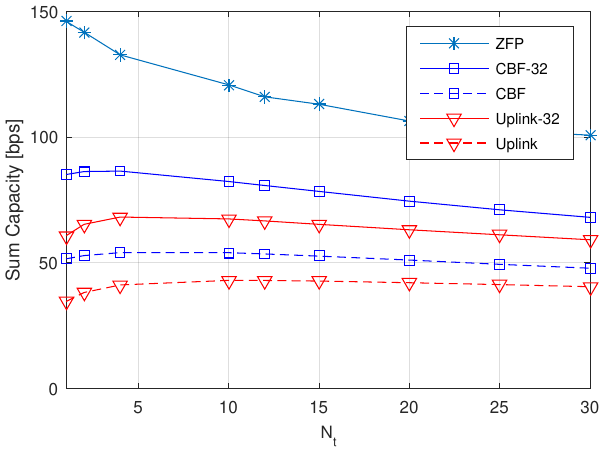}
\label{fig:result1} 
}
\hspace{0mm}
\subfloat[]{
\includegraphics[width=0.32\textwidth]{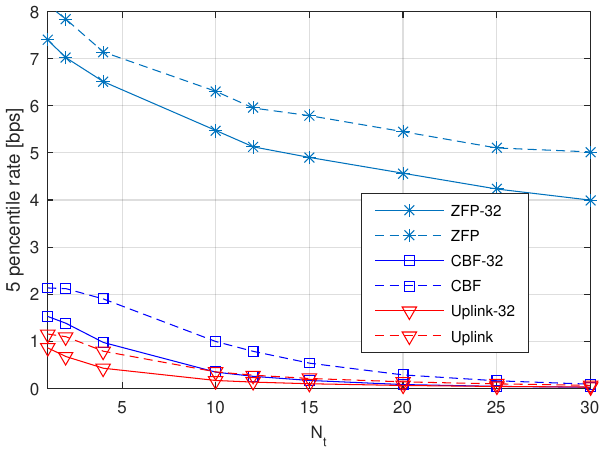}
\label{fig:result2} 
}
\hspace{0mm}
\subfloat[]{
\includegraphics[width=0.32\textwidth]{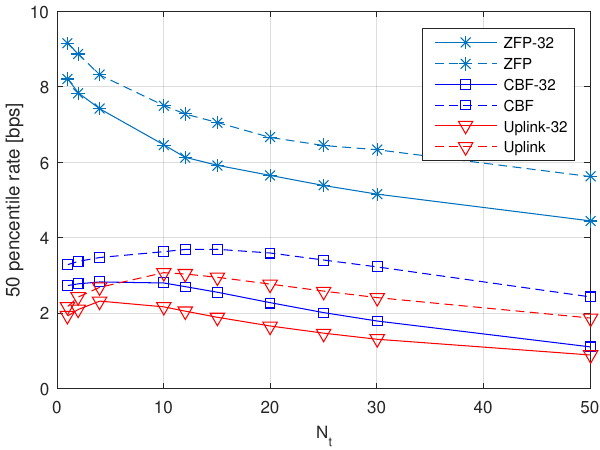}
\label{fig:result3}
}
}
\hspace{0mm}
 \caption{Performance of cell-free massive MIMO systems in terms of the number of antennas at each AP (i.e., $N_t$), where a total of $M=300$ antennas serve $K=16$ (default) or $K=32$ users (marked by -32 in the legend), including (a) the sum capacity; (b) the 5-percentile spectral efficiency; and (c) the 50-percentile spectral efficiency.    }
 \label{fig:result}
\end{figure*}

\subsubsection{Zero-Forcing Precoding}
Instead of maximizing the strength of the desired signal, ZFP aims to cancel IUI at the UE receiver \cite{Ref_jiang2022deep}. Its precoding matrix is the pseudo inverse of the channel matrix, i.e., $\textbf{B}=\hat{\textbf{G}}^*(\hat{\textbf{G}}^T\hat{\textbf{G}}^*)^{-1}\otimes \mathbf{E}$, where $\left[\textbf{E}\right]_{mk}=\sqrt{\eta_{mk}}$. As proved by \cite{Ref_nayebi2017precoding}, it is necessary to have $\eta_{1k}=\eta_{2k}=\cdots=\eta_{Mk}$, $\forall k$ to keep $\hat{\textbf{G}}^T\textbf{B}$ orthogonal such that the IUI is eliminated. Hence, we have $\eta_{mk}=\eta_{k}$, $\forall m$ and therefore $\textbf{B}=\hat{\textbf{G}}^*(\hat{\textbf{G}}^T\hat{\textbf{G}}^*)^{-1} \mathbf{D}$, where $\textbf{D} = \mathrm{diag}([\eta_1,\ldots,\eta_K])$. 
The IUI term in \eqref{EQN_downlinkModel} is zero-forced, namely $\sum_{i=1,i\neq k}^K \textbf{g}_k^T \textbf{b}_{i}  u_{i}=0$,  yielding $r_k =  \sqrt{p_d}\textbf{g}_k^T\textbf{b}_k u_k +w_k$, which is further derived to
\begin{align}\nonumber 
    r_k &=  \sqrt{p_d}\hat{\textbf{g}}_k^T\textbf{b}_k u_k +\sqrt{p_d}\tilde{\textbf{g}}_k^T\textbf{b}_k u_k+w_k\\
        &=  \sqrt{p_d \eta_k} u_k +\sqrt{p_d}\tilde{\textbf{g}}_k^T\textbf{b}_k u_k+w_k.
\end{align}
The effective SINR for user $k$ can be expressed as follows:
\begin{equation}  
    \gamma_k^{zfp} = \frac{p_d\eta_k}{\sigma_n^2 + p_d\sum_{i=1}^K \eta_i \chi_i^k}.
\end{equation}
where $\chi_{i}^k$ represents the $i^{th}$ diagonal element of the $K\times K$ matrix dedicated to user $k$ \cite{Ref_jiang2021impactcellfree}: $\mathbb{E}\biggl[ \left(\hat{\mathbf{G}}\hat{\mathbf{G}}^H\right)^{-1} \hat{\mathbf{G}}\mathbb{E}\left[\tilde{\textbf{g}}_k^H\tilde{\textbf{g}}_k\right]  \hat{\mathbf{G}}^H\left(\hat{\mathbf{G}}\hat{\mathbf{G}}^H\right)^{-1}\biggr]$,
and $\mathbb{E}\left[\tilde{\textbf{g}}_k^H\tilde{\textbf{g}}_k\right]$ is a diagonal matrix with the $k^{th}$ diagonal element equaling to $\beta_{mk}-\alpha_{mk}$. 
The cost-effectiveness of ZFP-based cell-free massive MIMO can be assessed through 
\begin{equation}
    \Gamma_{dl}^{zfp}=\frac{\sum_{k=1}^K \log(1+\gamma_{k}^{zfp})}{  N_{AP}( C_{f}+ N_{t} C_v )   }.
\end{equation}

\begin{figure*}[!tbph]
\centerline{
\subfloat[]{
\includegraphics[width=0.32\textwidth]{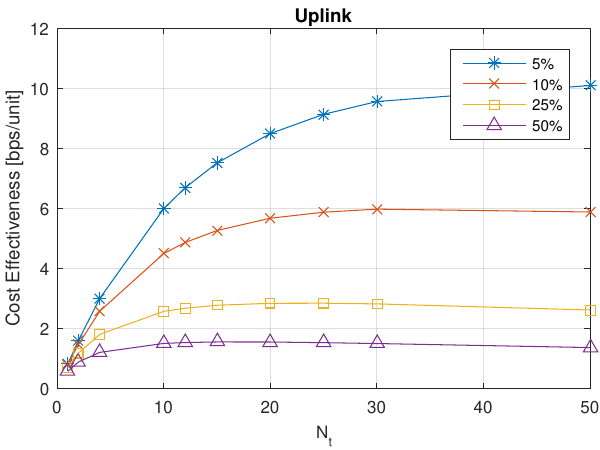}
\label{fig:ce1} 
}
\hspace{0mm}
\subfloat[]{
\includegraphics[width=0.32\textwidth]{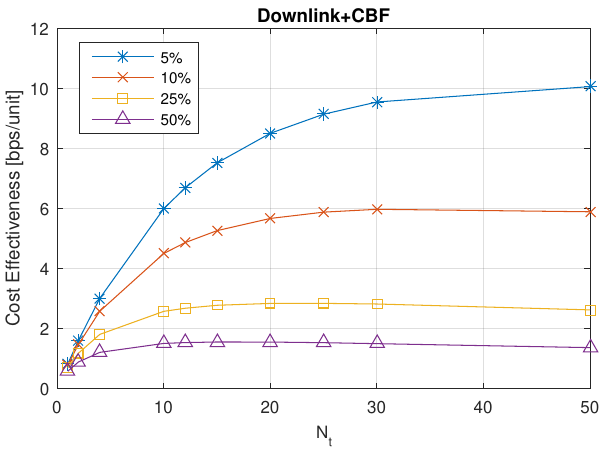}
\label{fig:ce2} 
}
\hspace{0mm}
\subfloat[]{
\includegraphics[width=0.32\textwidth]{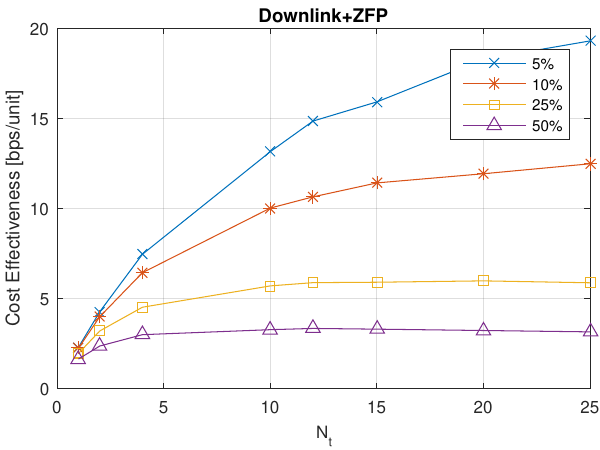}
\label{fig:ce3}
}
}
\hspace{0mm}
 \caption{Cost-effectiveness results of cell-free massive MIMO systems in terms of the number of antennas at each AP (i.e., $N_t$), where a total of $M=300$ antennas serve $K=16$. The ratio between the $N_t$-dependent and $N_t$-independent costs takes four values, i.e.,  $C_v/C_f=0.05,0.1,0.25$ and $0.5$.  }
 \label{fig:CEresult}
\end{figure*}

\section{Numerical Results}
The spectral efficiency and cost-effectiveness of a cell-free massive MIMO system are numerically evaluated in terms of the number of antennas per AP. 
Consider a configuration where a total of $M=300$ antennas serve users across a square area of $1\times 1\mathrm{km^2}$. The large-scale fading is figured out by the formula $\beta_{mk}=10^\frac{\mathcal{L}_{mk}+\mathcal{X}_{mk}}{10}$. The shadowing $\mathcal{X}_{mk}$, commonly modeled as a log-normal distribution $\mathcal{N}(0,\sigma_{sd}^2)$, where $\sigma_{sd}=8\mathrm{dB}$.  The path loss, determined by the COST-Hata model \cite{Ref_ngo2017cellfree}, is expressed as
\begin{equation} 
    \mathcal{L}_{mk}= \begin{cases}
-L_0-35\log_{10}(d_{mk}), &  d_{mk}>d_1 \\
-L_0-10\log_{10}(d_1^{1.5}d_{mk}^2), &  d_0<d_{mk}\leq d_1 \\
-L_0-10\log_{10}(d_1^{1.5}d_0^2), &  d_{mk}\leq d_0
\end{cases},
\end{equation}
where $d_{mk}$ is the distance between user $k$ and antenna $m$, the three-slope breakpoints are set as $d_0=10\mathrm{m}$ and $d_1=50\mathrm{m}$, and $L_0=140.72\mathrm{dB}$ is defined by 
\begin{IEEEeqnarray}{ll}
 L_0=46.3&+33.9\log_{10}\left(f_c\right)-13.82\log_{10}\left(h_{AP}\right)\\ \nonumber
 &-\left[1.1\log_{10}(f_c)-0.7\right]h_{UE}+1.56\log_{10}\left(f_c\right)-0.8
\end{IEEEeqnarray}
Here, the carrier frequency is $f_c=1.9\mathrm{GHz}$, the height of the AP antenna is $h_{AP}=15\mathrm{m}$, and the height of the UE is $h_{UE}=1.65\mathrm{m}$. 

The per-antenna and UE power constraints are $p_d=p_u=200\mathrm{mW}$. 
The white noise power density is $-174\mathrm{dBm/Hz}$ with a noise figure of $9\mathrm{dB}$, and the signal bandwidth is set at $5\mathrm{MHz}$. In the uplink, performed in a distributed manner, it is reasonable for each UE to adopt a full-power strategy with $\eta_k=1$. Regrettably, the optimal max-min power-control schemes in the downlink, employing both CBF and ZFP, are too computationally complex for practical implementation. In line with the recommendation by \cite{Ref_nayebi2017precoding}, we opt for sub-optimal schemes characterized by lower complexity. To elaborate, within the ZFP approach, we set $\eta_{1}=\ldots=\eta_{K}=\left( \max_m  \sum_{k=1}^{K} \delta_{km} \right)^{-1}$, where $\boldsymbol \delta_m= [\delta_{1m},\ldots,\delta_{Km}]^T=\mathrm{diag}(\mathbb{E}[  (\hat{\mathbf{G}}\hat{\mathbf{G}}^H)^{-1}    \hat{\mathbf{g}}_m \hat{\mathbf{g}}_m^H   \hat{\mathbf{G}}\hat{\mathbf{G}}^H)^{-1} ])$ and $\hat{\mathbf{g}}_m$ represents the $m^{th}$ column of $\hat{\mathbf{G}}$. In the case of CBF, the APs employ a full-power strategy, mathematically denoted as $\eta_{m}=(\sum_{k=1}^{K} \alpha_{mk} )^{-1}$, $\forall m$.

During the simulations, the number of AP antennas varies as $N_t\in\{1,2,4,10,12,15,20,25,30,50\}$. Correspondingly, a total of $300$ antennas were distributed to $N_{AP}\in\{300,150,75,30,25,20,15,10,6\}$ APs. Two groups of simulations were performed with the number of users set to $K=16$ and $K=32$, respectively. \figurename \ref{fig:result}a illustrates the sum rate for three scenarios: uplink, downlink with ZFP, and downlink with CBF. As expected, in the downlink, ZFP outperforms CBF significantly due to ZFP's utilization of global CSI to eliminate inter-user interference for all users. In contrast, CBF relies on only local CSI for precoding. The superiority of ZFP comes at the cost of high signaling overhead, as the CSI must be delivered via the fronthaul network. While both CBF in the downlink and MRC in the uplink share the approach of maximizing the desired signal, CBF's performance surpasses that of MRC. This is attributed to the higher power consumption in the downlink, amounting to $M\times p_t=\SI{60}{\watt}$, compared to the uplink's $K\times p_u=\SI{3.2}{\watt}$ or $\SI{6.4}{\watt}$. To provide a clear illustration, we omit the ZFP curve with $32$ users as it deviates significantly from the curves of CBF and MRC. As observed, the sum rate of ZFP monotonically decreases with the increasing number of AP antennas. However, in the case of CBF and MRC, optimal performance is achieved using multi-antenna APs at $N_t=4$.
 
The concept of user-experienced data rate, as defined by 3GPP, is rooted in the $5^{th}$ percentile point ($5\%$) of the cumulative distribution function of user throughput. This metric provides a meaningful measurement of perceived performance, particularly at the cell edge. \figurename \ref{fig:result}b presents the $5^{th}$ percentile per-user spectral efficiency, offering a glimpse into how the cell-edge performance varies with $N_t$. Simultaneously, \figurename \ref{fig:result}c displays the $50^{th}$ percentile, or median, per-user spectral efficiency. Observing \figurename \ref{fig:result}b, it becomes evident that the cell-edge performance is sensitive to reductions in AP density. However, a crucial insight emerges: the majority of performance levels can be maintained when the number of AP antennas remains moderate, specifically when $N_t<5$. In contrast, the median spectral efficiency demonstrates greater adaptability to multi-antenna AP scenarios. The peak performance for CBF and MRC occurs when there are approximately $N_t=10$ antennas. While the performance of ZFP exhibits a monotonically decreasing trend with the increasing number of AP antennas, it performs still well when $N_t$ is high. 

We evaluated cost-effectiveness based on the achievable spectral efficiency of each AP per cost unit, varying the number of antennas at each AP, where a total of $M=300$ antennas serve $K=16$. To maintain generality, we set the costs independent of $N_t$ to a fixed value of one cost unit during all simulations ($C_f=1$). Across specific deployment scenarios, the fee for adding an extra antenna and its associated RF chain varies from $5\%$ to $50\%$ of the fixed cost. In particular, the ratio between $N_t$-dependent and $N_t$-independent costs takes four values: $C_v/C_f=0.05, 0.1, 0.25$, and $0.5$. The results, as depicted in the \figurename \ref{fig:CEresult}, show a decrease as the cost of $C_v$ increases, reflecting an elevation in the total deployment cost without necessarily indicating a decline in performance. 

\section{Conclusions}
This paper investigated the cost-effectiveness aspect of cell-free massive MIMO through comprehensive cost and performance analysis. 
Factors like site acquisition, fiber connection, maintenance, and hardware costs were examined. The findings underscored the efficacy of multi-antenna APs in enhancing cost-effectiveness, albeit with a trade-off: reducing AP density may compromise per-user spectral efficiency and sum capacity. However, we demonstrated that the majority of performance levels can be maintained by determining the suitable number of antennas per AP.  

\bibliographystyle{IEEEtran}
\bibliography{IEEEabrv,Ref_PIMRC24}

\end{document}